\documentclass[pra,nofootinbib,twocolumn,showpacs,preprintnumbers,amsmath,amssymb,floatfix,amstex,superscriptaddress]{revtex4}
\usepackage{graphicx,graphics,wrapfig,rotating}      
\usepackage{dcolumn}                    	     
\usepackage{bm,fancybox}                	     
\usepackage{times,euscript,oldgerm}           %
\usepackage[german,english]{babel}	

\begin{document}

\title{Generalized Quantum Baker Maps as perturbations of a simple kernel}

\author{Leonardo Ermann}
\email[Email address: ]{ermann@tandar.cnea.gov.ar}
\affiliation{%
Departamento de F\'{\i}sica, Comisi\'{o}n Nacional de Energ\'{\i}a At\'{o}mica.
Avenida del Libertador 8250 (C1429BNP), Buenos Aires, Argentina.
}%
\affiliation{%
Departamento de F\'{\i}sica, FCEyN, UBA, Pabell\'{o}n $1$
Ciudad Universitaria, 1428 Buenos Aires, Argentina.
}%
\author{Marcos Saraceno}
\email[Email address: ]{saraceno@tandar.cnea.gov.ar}
\affiliation{%
Departamento de F\'{\i}sica, Comisi\'{o}n Nacional de Energ\'{\i}a At\'{o}mica.
Avenida del Libertador 8250 (C1429BNP), Buenos Aires, Argentina.
}%
\affiliation{%
Escuela de Ciencia y Tecnolog\'\i a, Universidad Nacional de San
Mart\'\i n. Alem 3901 (B1653HIM), Villa Ballester, 
Argentina.}

\date{\today}

\begin{abstract}
We present a broad family of quantum baker maps that generalize the proposal of Schack 
and Caves to any even Hilbert space with arbitrary boundary conditions. 
We identify a structure, common to all maps consisting of a simple kernel perturbed by diffraction effects. 
This ``essential'' baker's map has a 
different semiclassical limit and can be diagonalized analytically for 
Hilbert spaces spanned by qubits. In all cases this kernel provides an 
accurate approximation to the spectral properties - eigenvalues and 
eigenfunctions - of all the different quantizations. 

\end{abstract}

\maketitle

\section{Introduction}

The accurate description of pure eigenstates of chaotic systems is still beyond us. The
initial Berry-Voros ansatz \cite{Berryvoros} assuming a microcanonical ``uniformity'' on the energy shell
in accordance with Schmirelman theorem \cite{schnirelman} leaves space for weaker structures that lead to scarring 
by unstable periodic orbits \cite{Heller}.
The choice of basis is of course crucial in achieving a simple description. The standard choice for numerical 
calculations is to diagonalize the hamiltonian - or the map - in an integrable basis: coordinate, momentum, oscillator,
 plane wave, etc. In these bases the chaotic eigenfunctions have very large participation ratios, of the order of the 
 Hilbert space dimension, assimilating them to random eigenfunctions and excluding any perturbative description. Another 
 approach is to use a basis of ``quasi-modes'' constructed on short periodic orbits \cite{Vergini} , which has been used 
 extensively to describe the eigenstates of the Bunimovich stadium. Still another approach has been the study of the 
 distribution of zeroes in analytic representations. \cite{lebouef}

In this paper we show that the eigenfunctions and eigenvalues of the quantized baker's 
map - and a great variety of its generalizations - can be approximated by
 a relatively simple basis which diagonalizes a kind of ``essential'' baker's map. In this basis the eigenfunctions of
  all quantizations of the baker's map have very small participation ratios indicating that the basis vectors are 
  excellent approximations to the eigenfunctions. Moreover the eigenvalues are well reproduced in first order 
  perturbation theory. This basis can be analytically constructed for special Hilbert space dimensions, notably 
  the qubit case when $D=2^N$ \cite{lakshmin3}.
Our approach extends the pioneering work by Lakshminarayan \cite{lakshmin1,lakshmin2}, who was the 
first to 
realize that some eigenfunctions had a simple structure when looked upon differently in the Walsh-Hadamard basis.

The organization of the paper is as follows: in Section II we review the quantization of the baker's map
and propose a wide generalization of the Schack and Caves \cite{schackcaves} construction for qubits by
allowing arbitrary Hilbert space dimensions and Floquet angles. Utilizing a mixture of analytical and
circuit techniques we give a semiclassical interpretation of this construction in terms of 
semiquantum maps that quantize the baker's map iterates \cite{saravoros}. We
show that all these maps share a common structure as the product of two unitary kernels: a fixed one
$\mathfrak{B}$ common to all families and an almost diagonal one containing diffraction effects. 
In Sec. III we analyze spectral properties of this family of maps and show that $\mathfrak{B}$ provides an accurate basis
for the description of eigenfunctions and eigenvalues. 
 When the dimension is $D=2^N$ this
basis can be analytically constructed and is labeled by primitive binary strings.
For maps with antisymmetric boundary conditions this basis improve the Hadamard
representation discovered by Lakshminarayan \cite{lakshmin1}.

\section{Baker's map quantizations}

\subsection{Quantum baker maps}

The classical baker map is an example of an intuitive canonical transformation that can be expressed 
in terms of symbolic dynamics using the binary Bernoulli shift. The map is defined
in the unit square phase space $(q,p\in[0,1))$ by
\begin{eqnarray}
q_{i+1}&=&2q_{i}-[2q_{i}]\\
p_{i+1}&=&(p_{i}+[2q_{i}])/2
\end{eqnarray}
where $[~]$ denotes the integer part, and $i$ is the discrete time. This map is
area-preserving, and geometrically stretches the square by a factor of two in the $q$ direction,
squeezes by a factor of a half in the $p$ direction, and then stacks the right half onto the
left.

The map has a simple symbolic dynamics involving the binary expansions of the
coordinates, $q=0.\epsilon_{0}\epsilon_{1}\ldots=\sum_{k=0}^{\infty}\epsilon_{k}2^{-k-1}$ and
$p=0.\epsilon_{-1}\epsilon_{-2}\ldots=\sum_{k=1}^{\infty}\epsilon_{-k}2^{-k}$
$(\epsilon_{i}\in{0,1})$. One point of the phase space is represented by a bi-infinite symbolic string
\begin{equation}
(p,q)=\ldots\epsilon_{-2}\epsilon_{-1}
\bullet\epsilon_{0} \epsilon_{1}\epsilon_{2}\epsilon_{3}\ldots
\end{equation}
and the baker map action upon symbols is
\begin{equation}
(p,q)\longrightarrow (p',q')=\ldots\epsilon_{-2}\epsilon_{-1}
\epsilon_{0}\bullet\epsilon_{1}\epsilon_{2}\epsilon_{3}\ldots
\end{equation}

The classical map has two symmetries: {\em time-reversal} (T) reversing the direction of the flow 
and exchanging $p\leftrightarrows q$; and {\em parity} (R) exchanging $q\rightarrow 1-q$,
$p\rightarrow 1-p$, and bitwise logical NOT ($0\leftrightarrows 1$) upon symbols.\\

A quantization of the baker map is well known and proceeds by first setting up a Hilbert space with
appropriate boundary conditions. The phase space is made compact by imposing perodic boundary
conditions, thus turning it into a $2$--torus. The corresponding quantum structure is characterized by quasi--periodic 
boundary conditions

\begin{equation} \label{eq:flo}
\psi(q+1)=e^{i2\pi\kappa}\psi(q),\quad\quad
\tilde{\psi}(p+1)=e^{-i2\pi\eta}\tilde{\psi}(p)
\end{equation}
where  $2\pi\eta$ and
$2\pi\kappa$ are arbitrary Floquet angles and $\psi,\tilde\psi$ are Fourier transformed pairs. 
Solutions to (\ref{eq:flo}) only exist if $hD=1$ with $D$ integer, and they span a Hilbert space
$\mathcal{H}_{D}^{\eta,\kappa}$ having finite dimension $D$. The position and momentum
eingevectors are $|q_{j}\rangle$ and $|p_{j}\rangle$ with eigenvalues $q_{j}=(j+\eta)/D$ and 
$p_{k}=(k+\kappa)/D$ respectively $(j,k=0,\ldots,D-1)$. The vectors of each basis are orthonormal,
$\langle q_{j}|q_{k}\rangle=\langle p_{j}|p_{k}\rangle=\delta_{jk}$ and the two bases are related
via the finite Fourier transform with arbitrary Floquet angles,
\begin{equation}
(\hat{F}_{D}^{\eta,\kappa})_{kj}\equiv
\langle p_{k}|q_{j}\rangle=\frac{1}{\sqrt{D}}e^{-i\frac{2\pi}{D}(j+\eta)(k+\kappa)}
\end{equation}

The quantization of the baker map on an even-dimensional Hilbert space
 can be achieved \cite{voros} converting the most significant bit of position
in the most significant bit of momentum. The matrix of the map in mixed representation has the
form of two blocks with the finite Fourier transform of size $D/2$ in each one. In position
representation we have
\begin{equation}
B_{pos}^{\eta,\kappa}=\big(F_{D}^{\eta,\kappa}\big)^{-1}
\left(
\begin{array}{cc} F_\frac{D}{2}^{\eta,\kappa} &0
\\0 &F_\frac{D}{2}^{\eta,\kappa} \end{array} \right)
\end{equation}

\begin{figure}[htp!]
\begin{center}
\includegraphics[width=0.4\textwidth]{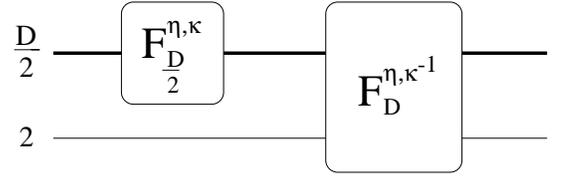}
\end{center}
\caption{\label{fig:bakerpos} Circuit representation of the Balazs-Voros-Saraceno quantum baker map. The thick line
represents a subsystem of dimension $D/2$.}
\end{figure}

This matrix product has a simple circuit representation in terms of the Fourier transform. 
This is shown in (Fig. \ref{fig:bakerpos}), where 
the lines represent subspaces ordered with the most significant one on the bottom, 
 the box is a unitary operator acting in the respective space, and the temporal flux is from left to
 right (opposite to the matrix representation) \cite{chuang}. 

The classical symmetries are preserved for some values of the Floquet angles. $T$ is preserved when 
$\eta=\kappa$, and $R$ when $\eta+\kappa=1$. 
The original quantization with $\eta=\kappa=0$ 
was made by Balazs and Voros \cite{voros}. 
In particular, the choice $\eta=\kappa=1/2$ (antisymmetric
boundary conditions) yields a map with both symmetries \cite{saraceno}.

We will use throughout techniques borrowed from quantum information theory 
\cite{chuang} allowing simple graphical representations of unitary operators in tensor product Hilbert spaces. 
For the baker's map the well known decomposition of the Fourier matrix into qubit operations \cite{Coppersmith} 
when $D=2^N$ allows a simple and efficient representation in terms of circuits. In Appendix A we show that this 
decomposition can be extended to arbitrary Floquet angles and other factorizations of $D$. 

\subsection{Family of baker maps on qubits}

The original quantization scheme only required $D$ to be even. In the special case when 
$D=2^{N}$, Schack and Caves proposed an entire class of quantum baker maps 
on $N$ qubits \cite{schackcaves, cavesscott}. They connected the binary
representation of the classical baker map to the qubit structure using the partial
Fourier transform, defined in the general case as
\begin{equation}\label{eq:FouParc}
\hat{G}_{n}^{\eta,\kappa}\equiv\hat{1}_{2^{n}}\otimes\hat{F}_{2^{N-n}}^{\eta,\kappa},\quad\quad
n=1,\ldots,N
\end{equation}
where $\hat{1}_{2^{n}}$ is the unit operator on the first $n$ qubits, and
$\hat{F}_{2^{N-n}}^{\eta,\kappa}$ is the
Fourier transform on the remaining qubits.
This transformation is used to define orthonormal basis states that are localized on the unit-square phase
space \cite{cavesscott}.

This operator has a straightforward representation in terms of quantum circuits as
shown in Fig.2. 

\begin{figure}[htp!]
\begin{center}
\includegraphics[width=0.4\textwidth]{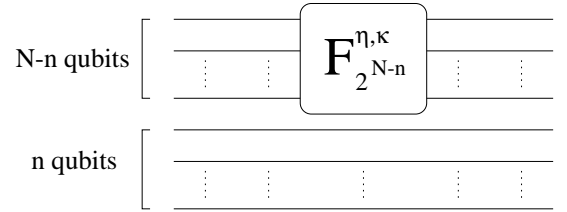}
\end{center}
\caption{\label{fig:FouParc} Circuit representation of the partial Fourier transform acting on the
N-n less significant qubits.}
\end{figure}

A quantum baker map with $N$ qubits is defined for each value of $n=1,\ldots,N$ by the unitary
operator \cite{schackcaves}
\begin{equation}\label{eq:FamBak}
\hat{B}_{N,n}\equiv
\hat{G}_{n-1}\circ\hat{S}_{n}\circ\hat{G}_{n}^{-1}
\end{equation}
where the dependence in $\eta$ and $\kappa$ is implicit. The shift operator 
$\hat{S}_{n}$ acts cyclically only on the first $n$ qubits, i.e., 
$\hat{S}_{n}|x_{1}\rangle |x_{2}\rangle \ldots |x_{n}\rangle
 |x_{n+1}\rangle \ldots |x_{N}\rangle=|x_{2}\rangle
 \ldots |x_{n}\rangle |x_{1}\rangle
 |x_{n+1}\rangle \ldots |x_{N}\rangle$.

\begin{figure}[htp!]
\begin{center}
\includegraphics[width=0.45\textwidth]{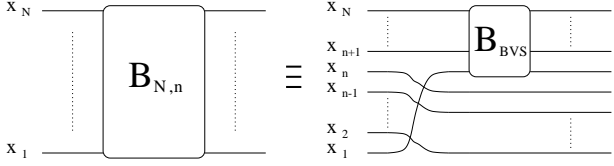}
\end{center}
\caption{\label{fig:cirfambak} Circuit representation of $\hat{B}_{N,n}$ in terms of the BVS
baker map}
\end{figure}

\begin{figure}[htp!]
\begin{center}
\includegraphics[width=0.5\textwidth]{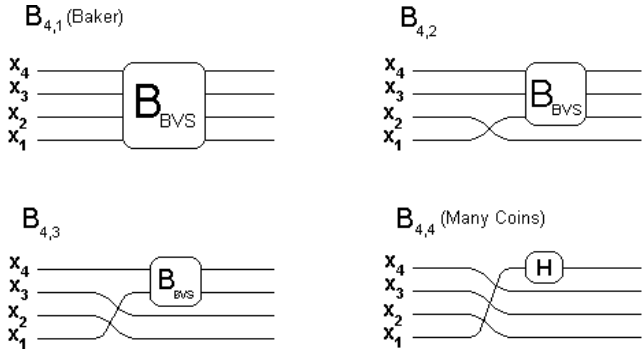}
\end{center}
\caption{\label{fig:cirfambak4Q} Circuit representation of the four members of the quantum 
baker map for four qubits ($\eta=\kappa=0$). $H$ is the Hadamard operator.}
\end{figure}

Since $\hat{S}_{1}$ is the unit
operator, $\hat{B}_{N,1}\equiv\hat{B}_{BVS}$ is the Balazs-Voros-Saraceno (BVS) map. On the other hand, 
it is easy to show that $\hat{B}_{N,N}$ is a map constructed only with swaps and one qubit Fourier transform. 
Since $\hat{S_n}$ commutes with $\hat{G}_{n}^{-1}$, $\hat{B}_{N,n}$, it can be written as
\begin{equation}\label{eq:FmBakCirc}
\hat{B}_{N,n}=\left( \hat{1}_{2^{n-1}}\otimes\hat{B}_{N-n+1,1}\right) \circ\hat{S}_{n}.
\end{equation}
Thus the action of $\hat{B}_{N,n}$ is equivalent to a cyclic shift of the $n$ most significant qubits followed by 
application of $B_{BVS}$ to the $N-n+1$ least significant qubits. This equivalence is shown
in circuit representation in Figs. \ref{fig:cirfambak} and \ref{fig:cirfambak4Q}.
All these maps have efficient implementations in terms of single qubit operations, swaps
and controlled phases, even for arbitrary Floquet angles. (see Appendix \ref{FouFloq}).

We will dedicate ourselves to $\hat{B}_{N,N}$, the last map of this family, in section D.
This map has appeared in different contexts. In the literature on quantum walks \cite{brun} it is called the ``many coins'' map and is used 
to study quantum walks with many independent coin throws. In the context of quantum graphs it represents a De Bruijn 
graph with complex phases on its edges \cite{tanner}. Nonnenmacher has observed that if the Fourier 
transform is replaced by the Walsh-Hadamard transform in the Balasz-Voros scheme
then $\hat{B}_{N,N}$  is the Walsh quantization of the baker's map and constitutes a ``toy model'' for the baker's map
\cite{nonnen1}. Taking this observation into account we can also 
think of the intermediate members of the family as mixed 
Fourier - Walsh-Hadamard quantizations. 

If one is willing to admit a BVS map in dimension $D=1$, which would amount to a trivial
inessential phase, then $\hat{S}_N$ - the simple cyclic shift on $N$-qubits - could also
be considered as an extreme member of this family. It was in fact a quantization of the
baker's map proposed by Penrose \cite{Penrose}, but never pursued.

\subsection{Quantization of iterated map}

We can give a different -semiclassical- interpretation to the Schack and Caves construction that,
besides clearly showing that all maps in the family are equivalent in the semiclassical limit
($h=\frac{1}{N}\rightarrow 0$), allows a generalization to Hilbert space dimensions of a much wider class.

\begin{figure}[htp!]
\begin{center}
\includegraphics[width=0.38\textwidth]{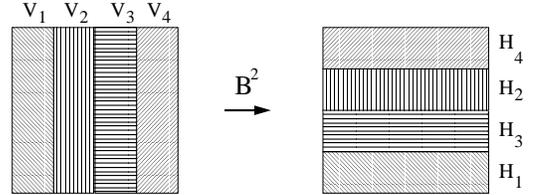}
\end{center}
\caption{\label{fig:BakClas2} Representation in phase space of two iterations of the classical baker map}
\end{figure}

We start by recalling that the scheme utilized by Balazs and Voros \cite{voros} to quantize the original baker map can also be 
utilized to quantize its iterates. We denote $B^{(T)}$ the unitary map
resulting from the direct quantization of the $T$ iteration of the classical map. Since the classical map divides 
the $q$ or $p$ segments into $2^{T}$ equal
intervals, we should initially assume that $2^{T}$ divides $D$ so that each strip is quantized by $D/2^T$ states. 
Notice that for this construction to be semiclassically plausible there should be many quantum states in each strip 
thus requiring $D/2^T \to \infty$.
As an example, two iterations of the classical baker map require four strips as in
 Fig. \ref{fig:BakClas2}.

The quantization of this second iterate proceeds exactly as for the first. The map 
$B^{(2)}$ can be quantized in the mixed representation as
\begin{equation}
B^{\eta,\kappa \ (2)}_{mix}= \left(
\begin{array}{cccc}F_{\frac{D}{4}}^{\eta,\kappa}&0&0&0\\0&0&F_{\frac{D}{4}}^{\eta,\kappa}&\\
0&F_{\frac{D}{4}}^{\eta,\kappa}&0&0\\0&0&0&F_{\frac{D}{4}}^{\eta,\kappa} \end{array} \right)
\end{equation}

These quantizations for various values of $T$ were studied in \cite{saravoros}, and
provide a {\em semiquantum dynamics}, somewhat intermediate between semiclassical and quantum propagation. 
The matrix representation of $B^{(T)}$ can be expressed easily using finite strings of bits in symbolic dynamics.
\begin{equation}
\langle m|B^{T}_{mix}|n\rangle=\delta_{\boldsymbol{\mu}^{\dag}\boldsymbol{\nu}}
\otimes F^{\eta\kappa}_{D/2^{T}},\ \ \boldsymbol{\nu}=\boldsymbol{\nu}(q_{n}),
\ \ \boldsymbol{\mu}=\boldsymbol{\mu}(p_{m})
\end{equation}
where
$\boldsymbol{\nu}(q_{0})=(\epsilon_{0}\ldots\epsilon_{T-1})=$ and 
$\boldsymbol{\nu}^{\dag}(q_{0})=(\epsilon_{T-1}\ldots\epsilon_{0})$.

\begin{figure}[htp!]
\begin{center}
\includegraphics[width=0.4\textwidth]{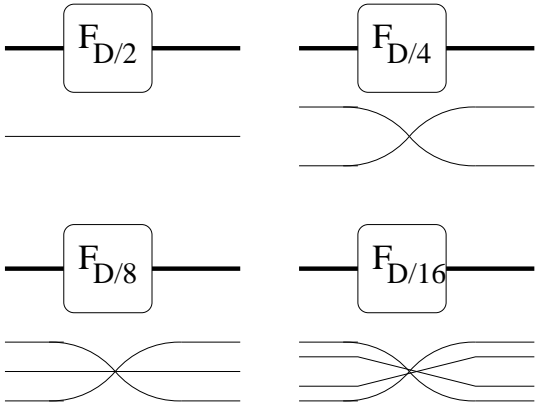}
\end{center}
\caption{\label{fig:Bakerequivmix} Circuits in the mixed representation for the quantization of the first 
four iterations of the baker's map.}
\end{figure}

The tensor product sctructure of $B^{(T)}$ allows a very simple circuit representation (Fig. 
\ref{fig:Bakerequivmix}), using the swap operator $\hat{S}_2$. The inversion of the order of the 
most significant qubits corresponds to the transfer of a finite string of $T$ bits of symbolic dynamics from 
momentum to coordinate and is accomplished by  $\frac{T}{2} (\frac{T-1}{2})$ swaps for even (odd) $T$.

It is important to observe that the direct quantization of the $T$-step classical map and the
$T$-th iteration of the one-step quantum map do not produce exactly the same matrices. 
In fact quantization and propagation are operations whose commutativity is only justified
in the semiclassical limit $D/2^T \to \infty$. The Schack and Caves
construction can then be seen from a different perspective using the fact that
$B^{(T+T')}$ and $B^{T}B^{T'}$ represent different matrices which are only asymptotically equal whenever 
the semiclassical condition is satisfied. Thus we can define families of quantizations of the one step map as
\begin{equation}\label{eq:defbalfam1}
\hat{B}_{D,2^{n-1}}\equiv\hat{B}^{(n-1)\dag}\hat{B}^{(n)} ~~~~~~~~~~~~n=1,...N
\end{equation}
where the only requisite is that $2^{N}$ divides $D$, and the BVS map is recovered for $\hat{B}_{D,1}$.
 It is clear from
the semiclassical arguments in \cite{saravoros} that all these families are semiclassically
equivalent in the sense that they all represent valid, and different, quantizations of the one--step
baker map. However, this equivalence requires that $n$ remain fixed while
$\frac{D}{2^{n}}\rightarrow\infty$. If, on the other hand we allow $2^{n}\sim D$ we can expect strong 
quantum deviations from this equivalence.

\begin{figure}[htp!]
\begin{center}
\includegraphics[width=0.48\textwidth]{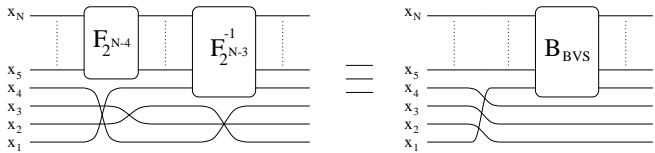}
\end{center}
\caption{\label{fig:bakerequiv}Equivalence of the quantum baker map with the semiclassical definition
and with the Schack and Caves maps.}
\end{figure}

If we work with a Hilbert space spanned by $N$ qubits and for antiperiodic boundary conditions, it is easy to see that 
the quantization obtained as a product of two semiquantum maps is equivalent to
the family  $\hat{B}_{N,n}$ of Eq.\ref{eq:FamBak}
obtained by Schack and Caves. 
In Fig. \ref{fig:bakerequiv}, for example, we use circuits to show that
 $\hat{B}^{(3)\dag}\hat{B}^{(4)}\equiv \hat{B}_{N,4}$.

In what follows we will use arbitrary Hilbert space dimensions. 
Therefore we will label the different maps not by the number of
qubits but by this dimension. Thus the Schack and Caves family will be labeled as $\hat{B}_{2^{N},2^{n-1}}$.

It should be clear that many more inequivalent constructions are posible along these lines, 
i.e. $B^{(3)}B^{(1)\dag}B^{(1)\dag}$ and many other similar combinations.

This interpretation of the family as a product of semiquantum maps gives also a natural explanation to the result
 of \cite{scott} who found that $B_{N,N}$ was abnormal in that
its classical limit was different from the baker map. Seen from a semiclassical point of view,
this result is quite natural as it would involve the quantization of strips of dimension
$D/2^N=1$ by means of rank one projection operators, thus strongly violating any semiclassical justification.
 However, as we show in Sec.III, it is precisely this ``extreme'' quantum map that provides the approximate description 
 of all the other quantizations.

\subsection{Quantum baker's map generalization}

Motivated by the Schack and Caves families for a space spanned by
qubits (Eq. \ref{eq:FamBak}),
 we will generalize the quantum baker's map (QBM) to Hilbert spaces with arbitrary Floquet angles and arbitrary even
dimension. Assume that $D=2 D_\alpha D_\beta$ in which case we interpret the map as operating
in a product Hilbert space $\mathcal{H}_{D}=\mathcal{H}_{2}\otimes\mathcal{H}_{D_\alpha}\otimes\mathcal{H}_{D_{\beta}}$. 
We relabel the coordinate states as
\begin{equation}
|j\rangle \to |\epsilon j_\beta j_\alpha\rangle=|\epsilon\rangle\otimes |j_\beta\rangle\otimes |j_\alpha\rangle
\end{equation}
where $j=\epsilon D_{\alpha} D_{\beta} + j_{\beta} D_{\alpha} +j_{\alpha}$ and $\epsilon=0,1$; 
\mbox{$0\le j_{\beta} < D_{\beta}$;} 
$0\le j_{\alpha} < D_{\alpha}$ thus making
$\mathcal{H}_{2}$ the ``most significant'' subspace and $\mathcal{H}_\alpha$ the
least significant one.
The QBM families are defined as
\begin{equation}\label{eq:defQBM}
\hat{B}_{D,D_{\beta}}=\left( \hat{1}_{D_{\beta}}\otimes\hat{B}\right) 
\circ\hat{S}_{2,D_{\beta}}
\end{equation}
where $\hat{B}$ is the BVS baker's map in a $2D_{\alpha}$-dimensional
Hilbert space with implicit dependence on the Floquet angles. 
The shift $\hat{S}_{2,D_\beta}$ between a qubit
and a $\beta$ subspace is a particular case of a shift defined as

\begin{equation}\label{eq:defshiftab}
\hat{S}_{D_{1},D_{2}}=\sum_{j=0}^{D-1}|D_1j-[\frac{j}{D_2}](D-1)\rangle~~\langle j|
\end{equation}
where $[j]$ is the integer part of $j$. This shift is a permutation of the states $j\in\mathcal{H}_{D}$ which exchanges the
significance of the $D_{1},D_{2}$ subspaces.

\begin{figure}[htp!]
\begin{center}
\includegraphics[width=0.4\textwidth]{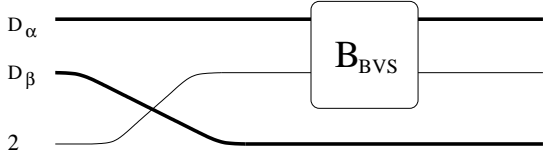}
\end{center}
\caption{\label{fig:FamDaDb} Circuit representation for the quantum baker map families. Each
line represents a subspace with the most significant one at the bottom.}
\end{figure}

$\hat{B}_{D,D_{\beta}}$ has a very simple circuit representation (Fig. \ref{fig:FamDaDb}) which obviously generalizes 
the qubit circuit of Fig. \ref{fig:bakerequiv}.\\ 
Clearly different maps
with a constant value of $D_\alpha D_\beta$ constitute a family of quantizations that generalize the previous qubit 
construction, which is recovered by choosing antiperiodic boundary conditions and a Hilbert space spanned by qubits.

We now arrive at the main point of our construction: the baker's map in Fig. \ref{fig:FamDaDb} can be
split, using the factorization properties of the discrete Fourier transform (cf. Fig. \ref{fig:bakersplit} in Appendix A) as
the product of two unitary kernels (Fig.\ref{fig:kernels}). 
\begin{figure}[htp!]
\begin{center}
\includegraphics[width=0.45\textwidth]{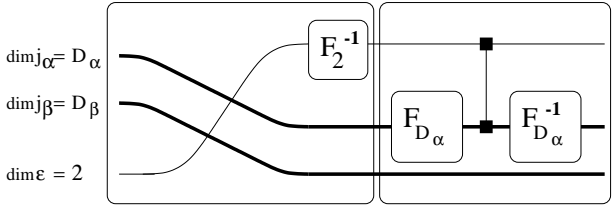}
\end{center}
\caption{\label{fig:kernels}  Circuit representation for the QBM family defined in Eq. 
\ref{eq:defQBM}. Each line represents a subspace and the squares respresent the phase interaction (see App. A).} 
\end{figure}
The first one which can be considered as an ``essential'' baker's map, is an obviuos
generalization of the ``many coins'' map of Fig. \ref{fig:cirfambak4Q}, and is common to all families.
We define it as
\begin{subequations}
\begin{eqnarray}\label{eq:essbak}
\hat{\mathfrak{B}} &=&e^{-i2\pi\eta\kappa} \ \hat{S}\circ\left(\hat{F}_{2}^{\dag}\otimes 
\hat{1}_{D_{\alpha}D_{\beta}}\right)\\
&=&e^{-i2\pi\eta\kappa} \left(\hat{1}_{D_{\alpha}D_{\beta}}\otimes \hat{F}_{2}^{\dag}\right)
\circ\hat{S}\label{eq:EQBM}
\end{eqnarray}
\end{subequations}
where 
\begin{equation}\label{eq:defshift}
\hat{S}\equiv\hat{S}_{2,D/2}
\end{equation}
is the shift operator used in \cite{lakshmin3,lakshmin1}.

The position representation of $\hat{\mathfrak{B}}$ is a square \mbox{$2D_{\alpha}D_{\beta}\times
2D_{\alpha}D_{\beta}$} complex matrix.
\begin{equation}
\hat{\mathfrak{B}}={e^{-i2\pi\eta\kappa}}
\begin{pmatrix} a&\ldots&0&
b&\ldots&0\\
c&\ldots&0&
d&\ldots&0\\
\vdots&\ddots&\vdots&\vdots&\ddots&\vdots\\
0&\ldots&a&0&\ldots&b\\
0&\ldots&c&0&\ldots&d 
\end{pmatrix}
\end{equation}
where
\begin{equation}
\hat{F}^{\dag}_{2}=
\begin{pmatrix}
a&b\\c&d
\end{pmatrix}
=\frac{e^{i\pi\eta\kappa}}{\sqrt{2}}
\begin{pmatrix}
1&e^{i\pi\kappa}\\e^{i\pi\eta}&e^{i\pi(\eta+\kappa+1)}
\end{pmatrix}
\end{equation}\\

The second kernel in Fig. \ref{fig:kernels} contains all the peculiarities of the different family members. It is diagonal in the subspace $\mathcal{H}_{\beta}$,
while the interaction of the qubit line with subspace $\mathcal{H}_{\alpha}$ gives raise to diffraction effects. The matrix elements of this
kernel are

\begin{equation}\label{eq:diffker1}
\langle\epsilon',j'_{\alpha},j'_{\beta}|\hat{K}_{D,D_{\beta}}| \epsilon,j_{\alpha},j_{\beta}\rangle=
\delta_{\epsilon,\epsilon'}\delta_{j_{\beta},j'_{\beta}}
\langle\epsilon,j'_{\alpha}|\hat{K}| \epsilon,j_{\alpha}\rangle
\end{equation}
where
\begin{subequations}\label{eq:diffker2}
\begin{equation}
\langle\epsilon,j'_{\alpha}|\hat{K}| \epsilon,j_{\alpha}\rangle =
e^{\frac{i\pi}{D_{\alpha}}\Phi}
\frac{\sin{(\pi D_{\alpha}\gamma)}}
{D_{\alpha}\sin{(\pi\gamma)}}
\end{equation}
with
\begin{eqnarray}
\Phi&=&
(j'_{\alpha}-j_{\alpha})(D_{\alpha}+2\kappa-1)+\nonumber\\
& &{}+(\epsilon-\eta)(\frac{1}{2}-\kappa)(D_{\alpha}-1)\\
\gamma&=&\frac{1}{D_{\alpha}}
\left(j'_{\alpha}-j_{\alpha}+\frac{\epsilon-\eta}{2}\right)
\end{eqnarray}
\end{subequations}

The strong forward diffraction peak gives this kernel an almost diagonal structure with weak off diagonal elements reflecting the block
structure of the different sizes of subspaces $\mathcal{H}_{\alpha}$, $\mathcal{H}_{\beta}$. 
When $D_{\alpha}=1$, the diffraction kernel is the identity, and for the qubit case this is the 
``many coins'' map \cite{brun,leo}. 

The generalizations of the QBM can be rewritten in terms of both kernels defined in Eqs. \ref{eq:essbak}, 
\ref{eq:diffker1} and 
\ref{eq:diffker2} as
\begin{equation}
\hat{B}_{D,D_{\beta}}=
(\hat{1}_{D_{\beta}}\otimes\hat{K})\circ\hat{\mathfrak{B}}
\end{equation}
where $\hat{B}_{D,1}\equiv \hat{B}$ is the $B_{BVS}$; and $\hat{B}_{D,D/2}\equiv\hat{\mathfrak{B}}$ 
is the {\em essential} quantum baker's map (EQBM).
A similar decomposition was obtained in \cite{soklakov} for the qubit case.

\begin{figure}[htp!]
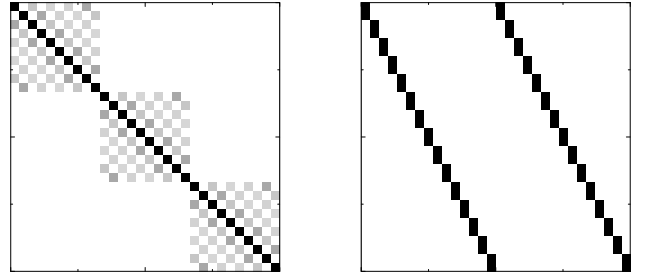

\begin{center}
\includegraphics[width=0.2\textwidth]{10.ps}\hspace{1cm}
\includegraphics[width=0.2\textwidth]{11.ps}
\end{center}
\caption{\label{fig:modKa5b3} Modulus of the matrix elements of the {\em diffraction} kernel (left) and
EQBM (right). $D=2D_{\alpha}D_{\beta}$ with $D_{\alpha}=5$, $D_{\beta}=3$ and the Floquet
angles $\eta=\kappa=0.5$ }
\end{figure}

The modulus of the matrix elements of both kernels for $D_{\alpha}=5$, $D_{\beta}=3$ are shown in Fig. \ref{fig:modKa5b3}.

It should be noticed that, even when antiperiodic boundary conditions are chosen, this family of quantizations does not respect the
time reversal symmetry of the original $B_{BVS}$ map. On the other hand they do commute with the parity $R$.\\

An alternative splitting of the QBM family can be considered if the $F^{-1}_{2}$ gate in Fig. \ref{fig:kernels} is included in the
diffraction kernel. In this case the resulting ``essential'' baker map is the shift map considered in \cite{lakshmin1}. 
We discuss the relative merits of these two splittings in the conclusions.

\section{Spectral properties}

\begin{figure}[htp!]
\begin{center}
\includegraphics[width=0.45\textwidth]{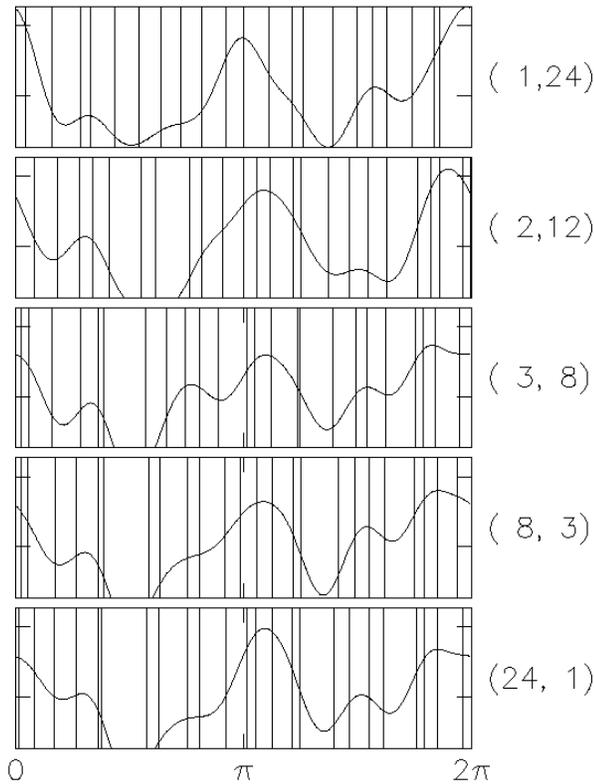}
\end{center}
\caption{\label{fig:EspFamBak48} Comparison  of the positive parity spectra of the family $B_{D,D_{\beta}}$ for $D=48$.
The values of $(D_{\alpha},D_{\beta})$ are on the right. The top spectrum corresponds to $\mathfrak{B}$ and the bottom 
to $B_{VBS}$. The continuous lines are the smoothed spectral densities.}
\end{figure}

Based on the fact that the diffraction kernel is almost diagonal we can attempt to describe the spectral properties of the family
$B_{D,D_{\beta}}$ in terms of those of the simpler $\mathfrak{B}$. That this is not a hopeless endeavour is shown in Fig.
\ref{fig:EspFamBak48} where we compare the positive parity spectra of the family with $D=48$ and various values of $D_{\alpha}$,
$D_{\beta}$. All quantizations share visually similar gaps and fluctuations indicating that a common structure is present. This fact
is even more apparent when the spectra are smoothed. The resulting spectral density is almost identical for all the maps.

\begin{figure}[htp!]
\begin{center}
\includegraphics[width=0.48\textwidth]{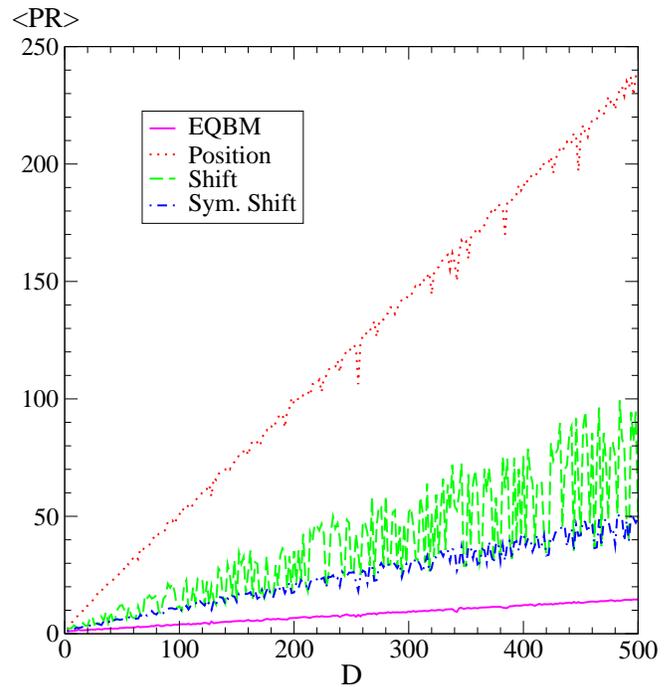}
\end{center}
\caption{\label{fig:PRD}(Color online) Participation ratio average as a function of the dimension for $B_{BVS}$ 
with antiperiodic boundary conditions in
different bases: position basis, shift basis, symmetrized shift basis and EQBM basis.}
\end{figure}

A more quantitative measure of the similarity of the eigenfunctions is given by the average participation ratio.
The participation ratio, $PR=(\sum_{r}|\langle \varphi_{r}|\psi\rangle|^{4})^{-1}$, is a rough measure of 
the number of basis elements $|\varphi_{r}\rangle$ needed to construct the state $|\psi\rangle$, with $PR\in[1,D]$. 
To assess the overall complexity of the eigenstates $|\Phi_{\alpha}\rangle$ in a given basis $|\varphi_{i}\rangle$ 
we compute the average PR
\begin{equation}
\langle PR\rangle=\frac{1}{D}\sum_{\alpha=0}^{D-1}\left(\sum_{i=0}^{D-1}|\langle\varphi_{i}|\Phi_{\alpha}\rangle|^{4}\right)^{-1}
\end{equation}
This quantity is plotted in Fig. \ref{fig:PRD} for $B_{BVS}$ eigenstates ($|\Phi_{\alpha}\rangle$) using several bases and
varying the dimension of the Hilbert space.

The $\langle PR\rangle$ in the position basis grows according to the random matrix theory predictions
(just below $D/2$) revealing the chaotic nature of the QBM. The $\langle PR\rangle$ decreases significantly using the shift 
eigenstates as a 
basis \cite{lakshmin1}, but has strong dependences on the dimension of the map which can be associated 
to periodic orbits of the map and number theory. The shift eigenstates are $|\chi_{\overline{\nu}}^{k}\rangle = \frac{1}{\sqrt{T}}
\sum_{j=0}^{T-1}e^{-i\frac{2\pi}{T}kj}|\nu_{j}\rangle$ where $|\nu_{j}\rangle$ are the trajectory points
$S^{j}_{2,D/2}|\nu_{0}\rangle$.

The symmetrized shift, used in \cite{lakshmin3}, is constructed as the parity 
projection of the shift basis, 
\begin{equation}
|\bar{\chi}_{\overline{\nu}}^{k}\rangle=\frac{1}{\sqrt{2}}\left(1\pm R\right)|\chi_{\overline{\nu}}^{k}\rangle
\end{equation}
where $R|j\rangle=|D-1-j\rangle$ is the parity acting in position basis.

This basis has a lower value of the $\langle PR\rangle$ and less fluctuations with the dimension $D$.
 In the case of EQBM's eigenstates as a basis, the $\langle PR\rangle$ is much lower and
grows slowly and smoothly with $D$.
The fact that $\langle PR\rangle$ is extremely small for the EQBM basis shows that it should be possible to describe the QBM
spectral properties (eigenfunctions and eigenvalues) in terms of the EQBM basis.

We have also explored the dependence of the growth of $\langle PR\rangle$ in the EQBM basis for different values of the Floquet parameters 
$((\eta,\kappa))$. Values close to the antisymmetric case $(1/2,1/2)$ yield consistently smaller values.

\subsection{The special case $D=2^N$.}

When $D=2^{N}$ the EQBM is the ``toy model'' considered in Refs. \cite{nonnen1,nonnen2}
 and its spectral properties can be analyzed exactly.
Beginning with the slightly more general operator $\hat{U}=\left(\hat{u}\otimes\hat{1}_{2^{N-1}}\right)\circ\hat{S}=\hat{S}\circ
\left(\hat{1}_{2^{N-1}}\otimes\hat{u}\right)$ where $\hat{S}$ is the shift operator defined in 
Eq. \ref{eq:defshift} and $\hat{u}$ is any unitary operator in a qubit space. The unitary matrix $\hat{u}$ can be 
diagonalized as 
\begin{equation}\label{eq:udiag}
u=A\left( \begin{array}{cc} e^{i\theta_{0}} & 0\\ 0 & e^{i\theta_{1}} \end{array} \right)A^{\dag}
\end{equation}
where $A$ is a unitary $2\times 2$ matrix. thus, the unitary matrix $U$ can be written in a 
simpler way as \mbox{$U=A^{\otimes N}U_{0}A^{\dag \otimes N}$ with $U_{0}$} in computational matrix representation as
\begin{eqnarray}
\langle j_{1}| U_{0}|j_{0} \rangle&=&\exp{ \left(i\theta_{0} +i \left[\frac{j_{0}}{2^{N-1}}\right]
(\theta_{1}-\theta_{0})\right)}\times\nonumber\\
& &\times\delta\left(j_{1}-2j_{0}+\left[\frac{j_{0}}{2^{N-1}}\right]
(2^{N}-1) \right)
\end{eqnarray}
This operator takes a very simple form if we label the states by a binary string, $|j\rangle\rightarrow|\nu\rangle\equiv
|a_{0}\ldots a_{N-1}\rangle$ with \mbox{$j=\sum_{i=0}^{N-1}a_{i}2^{N-1-i}$}. In that representation
\begin{equation}
U_{0}|a_{0}\ldots a_{N-2}a_{N-1}\rangle=e^{i a_{0}\theta_{1}+i(1-a_{0})\theta_{0}}|a_{1}\ldots a_{N-1}a_{0}\rangle
\end{equation} 
This is a permutation which will have cycles whose period $T_{\overline{\nu}}$ will depend on the 
binary structure of $\nu$. We then have 
\begin{equation}
U_{0}^{T_{\overline{\nu}}}|\nu\rangle=e^{i\Phi_{\overline{\nu}}}|\nu\rangle
\end{equation}
where $\Phi_{\overline{\nu}}=N_{0}\theta_{0}+N_{1}\theta_{1}$; $T_{\overline{\nu}}$ 
is the primitive period of the string $\nu$ and $N_{0}$
and $N_{1}$ are the number of $0$'s and $1$'s in the primitive string. Notice that $N_{0}+N_{1}=T_{\overline{\nu}}$ and that 
$T_{\overline{\nu}}$
must divide $N$. The eigenvalues of $U_{0}$ are then the shifted roots of unity
\begin{equation}\label{eigenvalphi}
\lambda_{\overline{\nu},k}=e^{i\frac{\Phi_{\overline{\nu}}+2\pi k}{T_{\overline{\nu}}}}
\end{equation}
and the corresponding eigenstates are
\begin{subequations}\label{eq:Psi0}
\begin{equation}
|\widetilde{\Psi}_{\overline{\nu}}^{k}\rangle=\frac{1}{\sqrt{T_{\overline{\nu}}}}\sum_{m=0}^{T_{\overline{\nu}}-1}
e^{i\varepsilon_{m}}{S}^{m}|\nu_{0}\rangle
\end{equation}
where
\begin{equation}
\varepsilon_{m}=m\left( \theta_{0}-\frac{\Phi_{\overline{\nu}}+2\pi k}{T_{\overline{\nu}}}\right)
+(\theta_{1}-\theta_{0})\sum_{i=0}^{m-1}a_{i}
\end{equation}
\end{subequations}
For example the primitive orbits for $N=4$ are $\overline{0}$, $\overline{1}$, $\overline{01}$, $\overline{0011}$,
$\overline{0001}$ and $\overline{0111}$ leading to two fixed points, one cycle of period 2 and 3 cycles of period 4.\\

The eigenstates of the original operator $U$ are then obtained as
\begin{equation}\label{eq:PsiPhi}
|\Psi_{\overline{\nu}}^{k}\rangle=A^{\otimes N}|\widetilde{\Psi}_{\overline{\nu}}^{k}\rangle
\end{equation}

When $u=e^{-i2\pi\eta\kappa}F^{\dag\ (\eta,\kappa)}_{2}$ its eigenvalues and eigenfunctions are given explicitly as
\begin{eqnarray}\label{avlF2}
\vartheta_{0,1}&=&e^{i\theta_{0,1}}=\pm\exp{\left(i\frac{\pi}{2}(\eta+\kappa-2\eta\kappa)\mp 
i\omega\right)}\\ \label{avcF2}
|\psi_{0,1}\rangle&=&\frac{1}{\sqrt{2}}\left[ 2\mp\sqrt{2} \cos{\left(\frac{\pi}{2}(\eta+\kappa)\mp\omega\right)}
\right]^{-\frac{1}{2}}\times\nonumber\\
& &\hspace{-1cm}\times\left[e^{i\pi\kappa}|0\rangle+(\pm\sqrt{2}e^{i\frac{\pi}{2}(\eta+\kappa)
\mp i \omega}-1)|1\rangle \right]
\end{eqnarray}
where $\omega\in[0,\pi/4]$ is defined as
\begin{equation}
\sin{(\omega)}=\frac{1}{\sqrt{2}}\sin{\left(\frac{\pi}{2}(\eta+\kappa)\right)}
\end{equation}
These expressions then provide the spectral properties of the EQBM in the qubit case $D=2^{N}$ for arbitrary Floquet
angles. 

For special values of $\eta, \kappa$ the matrix $F^{\dag\ (\eta,\kappa)}_{2}$ has a definite short
period, i.e. $\left(F^{\dag\ (0,0)}_{2}\right)^{2}=1$ and $\left(F^{\dag\ (0.5,0.5)}_{2}\right)^{4}=1$ thus leading to
a periodicity $2N$ or $4N$ for the full EQBM map. These short periodicities imply that the spectrum given by Eq. 
\ref{eigenvalphi}
will be highly degenerate for large values of $N$. Therefore the eigenfunctions in Eq. \ref{eq:Psi0} will not be unique and other
linear superpositions can also be constructed. They have been considered in \cite{nonnen2} and shown to have a multifractal
structure in the large $N$ limit. For others values of $\eta, \kappa$ the spectrum is in general not degenerate and this
possibility does not exist.

Notice that the eigenstates of $-i F^{\dag\ (0.5,0.5)}_{2}$ (with antiperiodic boundary
conditions) in Eq. \ref{avcF2} are $H|0\rangle$ and $H|1\rangle$
where $H$ is the Hadamard operator
\begin{equation}
H=\frac{1}{\sqrt{2}}\left( \begin{array}{cc} 1&1\\1&-1 \end{array}\right)
\end{equation}
Therefore, for this case $A=H$,  $\theta_{0}=\pi/2$ and $\theta_{1}=0$.
The explicit expression of EQBM eigenstates with their respective eigenvalues for 3 qubits are:
\begin{subequations}
\begin{eqnarray}
|\Psi_{\overline{0}}^{0}\rangle &=& H^{\otimes3}|000\rangle \ \ \ \ \ \  \longrightarrow 
\ \ \ \ \ \ \lambda_{\overline{0},0}=1 \nonumber\\
|\Psi_{\overline{1}}^{0}\rangle &=& H^{\otimes3}|111\rangle \ \ \ \ \ \  \longrightarrow 
\ \ \ \ \ \ \lambda_{\overline{1},0}=-i \nonumber\\
|\Psi_{\overline{001}}^{k}\rangle &=& \frac{1}{\sqrt{3}}H^{\otimes3}\big( |001\rangle +e^{i\pi\frac{1-4k}{6}}|010\rangle 
+ e^{i\pi\frac{1-4k}{3}}|001\rangle \big)\nonumber\\
& &\ \ \ \ \ \ \longrightarrow \ \ \ \ \ \ \lambda_{\overline{001},k}=e^{i\pi \frac{4k-1}{6}} \nonumber\\
|\Psi_{\overline{011}}^{k}\rangle &=& \frac{1}{\sqrt{3}}H^{\otimes3}\big( |011\rangle +e^{i\pi\frac{1-2k}{3}}|110\rangle 
+ e^{i\pi\frac{1-8k}{6}}|101\rangle \big)\nonumber\\
& &\ \ \ \ \ \ \longrightarrow \ \ \ \ \ \ \lambda_{\overline{011},k}=e^{i\pi \frac{2k-1}{3}} \nonumber
\end{eqnarray}
\end{subequations}
with $k=0,1,2$.\\

Notice that in this argument the role of the tensor product of Hadamard gates is that of diagonalizing 
$F^{\dag\ (0.5,0.5)}_{2}$. This is a peculiar case of the antisymmetric quantization ($\eta,\kappa=1/2,1/2$). Other values
would still lead to a tensor product of one--qubit gates but with a more complicated structure given by Eq.
\ref{avcF2}. This is different from the use of the Hadamard transform in \cite{lakshmin3} where its role is to restore
the parity symmetry to the shift map and is thus independent of the Floquet angles considered.

\subsection{Matrix of overlaps}

While the average PR gives a global measure of the similarity between the eigenbases of $B_{BVS}$
($|\Phi_{\alpha}\rangle$) and $\mathfrak{B}$ ($|\Psi_{\beta}\rangle$) a more detailed view is given by the matrix
of overlaps $M_{\beta \alpha}=|\langle\Psi_{\beta}|\Phi_{\alpha}\rangle|$.
If the eigenstates are ordered by their phases on the unit circle, the resulting matrix should be almost diagonal
with the off--diagonal elements signaling the importance of small components other than the main ones.

\begin{figure}[htp!]
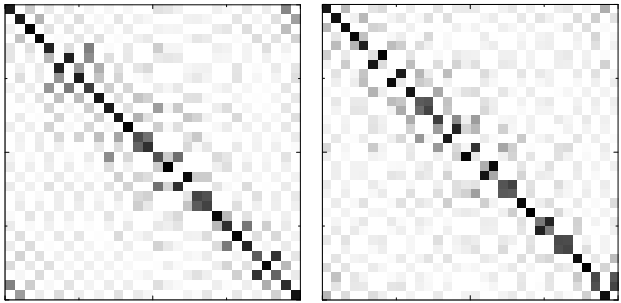

\begin{center}
\includegraphics[width=0.22\textwidth]{14.ps}\hspace{0.2cm}
\includegraphics[width=0.22\textwidth]{15.ps}
\end{center}
\caption{\label{fig:Prod} Matrix of overlaps $(M)_{\beta,\alpha}=|\langle\Psi_{\beta}|\Phi_{\alpha}\rangle|$ with 
Floquet angles $\eta=\kappa=0.5$ for $D=30$ (left) and $D=32$ (right). The 
eigenstates of both bases are ordered by growing eigenphases.}
\end{figure}

Fig. \ref{fig:Prod} shows the matrix of overlaps for $D=30$ and for the qubit case $D=32$. For the latter the spectrum
is concentrated at the values $e^{i2\pi\frac{k}{20}}$ and therefore many eigenstates are degenerate. In this case we
choose the basis as given in Eqs. \ref{eq:PsiPhi} and \ref{eq:Psi0}. For $D=30$ on the other hand there are no
degeneracies and the basis is unique. Fig. \ref{fig:Prod} shows clearly that in both cases most states
$|\Phi_{\alpha}\rangle$ have a very large overlap with one basis state (this overlap being typically $\sim0.9$) while a
few are mixtures of two. Some inversions in the ordering on the unit circle are also evident.\\

\begin{figure}[htp!]
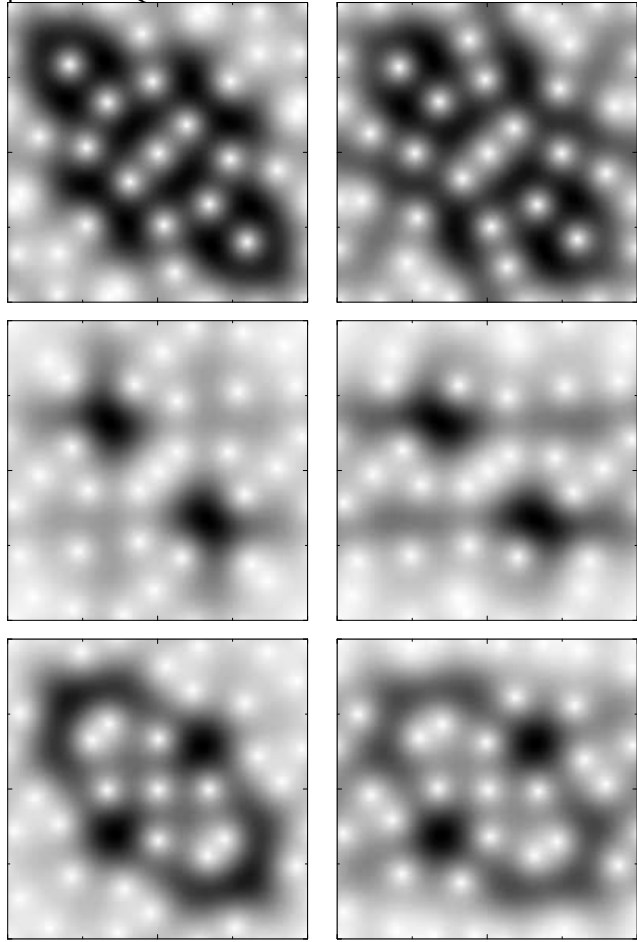

\begin{center}
\includegraphics[width=4cm]{16.ps}\hspace{0.3cm}
\includegraphics[width=4cm]{17.ps}\\ \vspace{0.2cm}
\includegraphics[width=4cm]{18.ps}\hspace{0.3cm}
\includegraphics[width=4cm]{19.ps}\\ \vspace{0.2cm}
\includegraphics[width=4cm]{20.ps}\hspace{0.3cm}
\includegraphics[width=4cm]{21.ps}
\end{center}
\caption{\label{fig:Hus} Husimi representation for three eigenstates of the $B_{BVS}$ (left) $|\Phi_{\alpha}\rangle$
and  $\mathfrak{B}$ ($|\Psi_{\beta}\rangle$) (right) in a Hilbert space spanned by 5 qubits.}
\end{figure}

The eigenfunctions are well approximated by the states that show no mixing in the matrix of overlaps. Fig. \ref{fig:Hus} shows the comparison of three
such eigenstates for $D=32$ in the Husimi representation.
The main difference in these eigenfunctions are due to the lack of time reversal symmetry of the basis states. This
symmetry, present in $|\Phi_{\alpha}\rangle$, is reflected clearly in the pattern of zeroes which is symmetric with
respect to the main diagonal.
Notice that, while the basis states for $D=2^{N}$ can be analytically constructed, the basis of the EQBM for
$D\neq 2^{N}$ has to be found numerically. What we show here is that EQBM provides an excellent starting point to study
the spectral properties of any baker quantization for all values of $D$.

\subsection{The spectrum}
Every QBM different from the EQBM has a complex spectrum which follows the RMT predictions 
(non-degenerate, with eigenvalues repulsion, etc. \cite{haake}). In general, the spectrum of EQBM will be 
``chaotic'' too. But, when the
Hilbert space is spanned by qubits and for special Floquet angles the EQBM has a short period and its spectrum is
degenerate.
Therefore, in these cases, the zeroth order approximation with EQBM's spectrum is not enough to accurately represent 
the spectrum of QBMs.

The first order aproximation can be computed multiplying the zeroth order
eigenvalues with the diagonal of the perturbation matrix (the diffraction kernel) 
represented in the EQBM basis. The corrected eigenvalues are then
\begin{equation}\label{eq:lambda0}
\lambda^{(1)}_{j}=\lambda^{(0)}_{j}\langle \Psi_{j}|K_{D,D_{\alpha}}| \Psi_{j} \rangle
\end{equation} 

We find that the diagonal elements of the diffraction kernel in Eq. \ref{eq:lambda0} are close to the unit circle, their
modulus tipically $\sim 0.9$. We adopt their phase as first order correction to the $\lambda^{(0)}_{j}$ eigenphases.

This approximation will be specially useful for the case of Hilbert spaces spanned by qubits since the corrections 
break degeneracy of the analytic EQBM's spectrum. Using Eqs. \ref{eq:diffker1}, \ref{eq:diffker2} and
\ref{eq:Psi0} the first
order approximation is 
\begin{widetext}
\begin{subequations}\label{eq:lambda1}
\begin{eqnarray}
\lambda_{\overline{\nu},k}^{(1)}&=&\frac{1}{T}\sum_{l,m=0}^{T-1}e^{i\xi(l,m)}\langle\nu_{0}|S^{\dagger l} 
A^{\dagger\otimes N} K_{D,D_{\alpha}} A^{\otimes N} S^{m}|\nu_{0}\rangle\\
\xi(l,m)&=&\theta_{0}(m-l)+(\theta_{1}-\theta_{0})\left( \sum_{i=0}^{m-1}a_{i}-\sum_{j=0}^{l-1}a_{j} \right)+
\frac{\phi_{\overline{\nu}}+2\pi k}{T}(l-m+1)
\end{eqnarray}
\end{subequations}
\end{widetext}

\begin{figure}[htp!]
\begin{center}
\hspace{-1.5cm}\includegraphics[width=10.1cm]{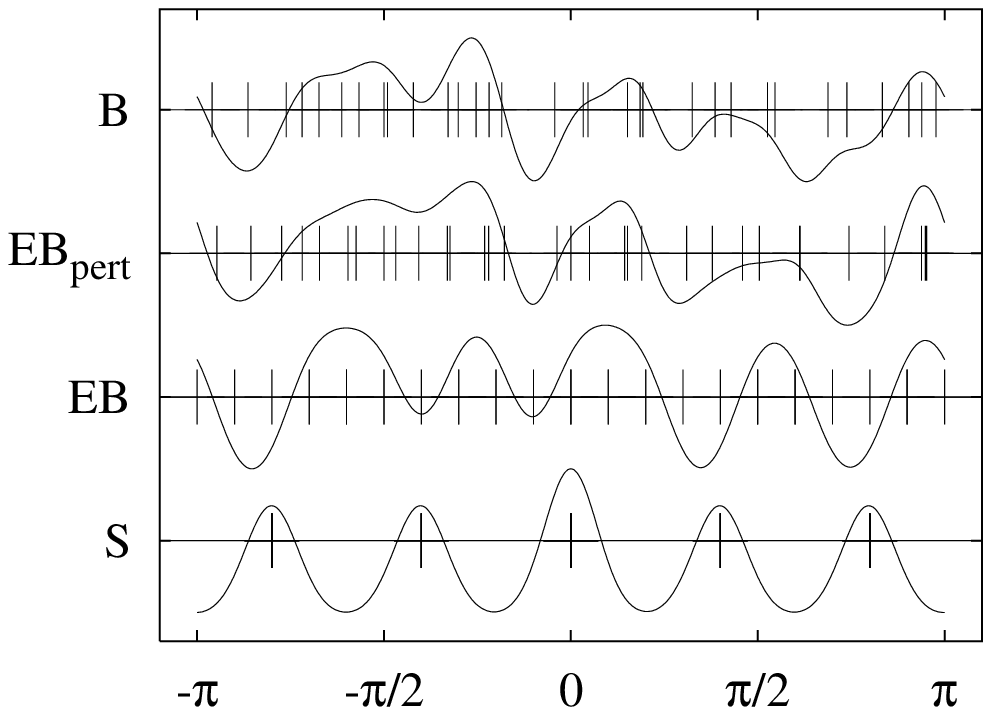}\\
\vspace{-0.5cm}
\hspace{-1.5cm}\includegraphics[width=10.1cm]{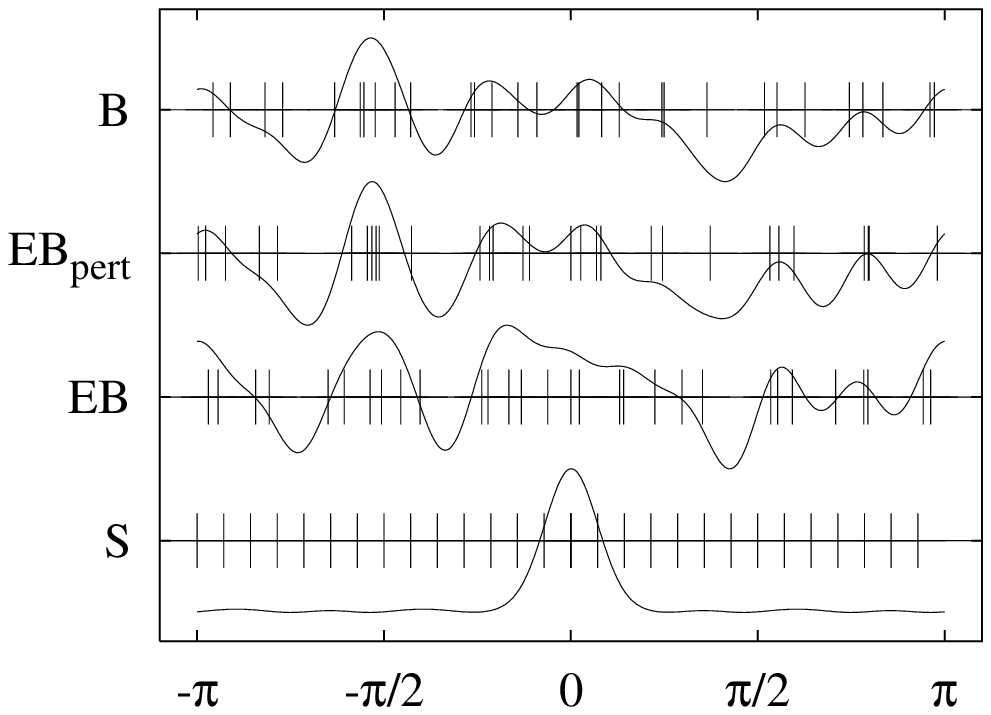}\vspace{-0.5cm}
\end{center}
\caption{ \label{fig:spectrum} Eigenphases of the shift operator ($S$), the EQBM ($EB$), 
the first order correction of the EQBM with the Kernel B ($EB_{pert}$), and the BVS Baker ($B$) for
dimension $D=32$ (top) and $D=30$ (bottom). The continuous lines are the smoothed spectral densities.}
\end{figure}

In Fig. \ref{fig:spectrum} we compare the spectra obtained with this approximation in the qubit case $D=32$
and for $D=30$.

In the non--qubit case ($D=30$) the EQBM spectrum is already quite accurate and is notably improved by the perturbation. It is
clear that the gaps and quasi degeneracies of the BVS are well reproduced and the smoothed density of eigenvalues is almost
identical. The qubit case is not as well approximated since the zeroth order is very degenerate and the corrections are too small. However it
is obvious that the smoothed spectrum is greatly improved by the correction.

\section{Conclusion}

We have shown that many valid quantizations of the baker's map have a simple matrix ($\mathfrak{B}$) as an essential
core.  This matrix
has a structure that captures many features of the complexity of eigenfunctions of these maps. In spite of its
simplicity, this matrix generally cannot be analytically diagonalized.
The numerical eigenfunctions are excellent approximations to most baker map's eigenfunctions with an \emph{average} participation
ratio of about $15$ at $D=500$. Many individual eigenfunctions have participation ratios of values near $1$.
The eigenvalues are also well approximated in their fluctuation properties. The smoothed density of states is
almost identical for all quantized families.
The case $D=2^{N}$ is very special. Some of its properties have been analyzed before as a toy model \cite{nonnen1} or as the extreme
case of the Schack and Caves family \cite{scott}.
Its spectrum is highly degenerate and its semiclassical limit is different from all of the others. Both the eigenvalues and the
eigenfunctions can be analytically obtained and the basis constructed on primitive binary cycles. They have participation ratios that
smoothly interpolate those of more general values of $D$.
An important ingredient in $\mathfrak{B}$ is the shift $\hat{S}$. In fact, we could well have shifted the $F_{2}^{\dag}$ gate in Fig.
\ref{fig:kernels} to the diffraction kernel and obtained a splitting in terms of the shift multiplied by a modified ``perturbation''. 
This procedure would yield as an ``essential'' baker just the shift $\hat{S}$ and would leave all the peculiarities of the
different families to a modified diffraction kernel. The shift basis, parity projected, was recently adopted in ref.
\cite{lakshmin3}. It has the advantage, over the EQBM basis, that it can be analytically constructed for all values of $D$
using the number theory properties of the shift. However, as we show in Fig. \ref{fig:PRD}, it consistently gives 
higher values participation ratio.

\begin{acknowledgments}

This work was financially supported by CONICET and the ANPCyT.

\end{acknowledgments}

\appendix
\section{Factorization of DFT with Floquet angles}\label{FouFloq} 

In the text we repeatedly used a circuit representation for the Fourier transform. 
Here we provide the details of this representation. The discrete Fourier transform in a $D$-dimensional Hilbert 
space with Floquet angles is
\begin{equation}\label{eq:Fou}
\langle j|F_D^{\eta,\kappa}|k\rangle=\frac{1}{\sqrt{D}}\exp[-i\frac{2\pi}{D}(j+\eta)(k+\kappa)]
\end{equation}

If $D=D_{1}D_{2}$, we can relabel the states as $j=j_{1}+D_{1}j_{2}$ and 
$k=D_{2}k_{1}+k_{2}$; with $0\leq j_{1},k_{1}\leq D_{1}-1$, and $0\leq j_{2},k_{2}\leq D_{2}-1$. Notice that 
index $j_2$ is more significant than $j_1$ and viceversa for $k_i$.
We then obtain

\begin{equation}\label{eq:DecompFou}
\langle j|F_{D_1D_2}^{\eta,\kappa}|k\rangle=
\langle j_1|F_{D_1}^{\eta,\kappa}|k_1\rangle
\langle j_2|F_{D_2}^{\eta,\kappa}|k_2\rangle \Theta(j_1,k_2)
\end{equation}
where, besides the Fourier transforms in each subspace, there is an 
an interaction term between $j_{1}$ and $k_{2}$ given by
\begin{equation}\label{eq:phaseint}
\Theta(j_1,k_2)=e^{i2\pi\eta\kappa} e^{-i2\pi\left(\frac{j_{1}+ \eta }{D_{1}}-\eta \right)
\left( \frac{k_{2}+\kappa}{D_{2}}-\kappa \right)}
\end{equation}

\begin{figure}[htp!]
\begin{center}
\includegraphics[width=0.45\textwidth]{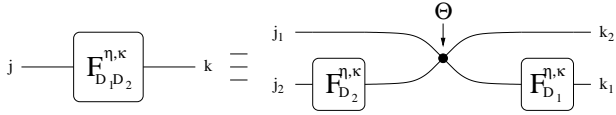}
\end{center}
\caption{\label{fig:FactorFou} Circuit representation of the factorization of the quantum Fourier
transform with Floquet angles in a Hilbert space with dimension $D=D_{1}D_{2}$}
\end{figure}

The circuit of this factorization is represented in Fig. \ref{fig:FactorFou}. 

The swap between the two subspaces is needed because of the different significance assigned 
to the $D_1,D_2$ subspaces in the decomposition of $j$ and $k$. As a matrix in the product
space this swap takes the permutation form
\begin{equation}
\hat{S}_{D_1D_2}=\sum_{j=0}^{D-1}|D_1j-\left[\frac{j}{D_2}\right](D-1)\rangle~~\langle j|
\end{equation}
where $[x]$ is the integer part. Notice that $\hat{S}_{D_2D_1}=\hat{S}_{D_1D_2}^{-1}$
For $D_1=2, \hat{S}_{2D_2}$ it is the shift operator utilized by Lakshminarayan \cite{lakshmin2}.

Clearly the factorization is recursive so that it can be iterated until the different prime factors of $D$ are reached. 
When $D=2^N$ and $\eta,\kappa=0$ the decomposition leads to the well known quantum circuit for the 
DFT \cite{Coppersmith} in terms of Hadamard gates and
diagonal phases. 

We give for reference the form of the interaction $\Theta$ for the special case of interest here of one qubit
interacting with a subspace of dimension $D$ ($D_{1}=2$, $D_{2}=D$) for the special values $\eta,\kappa=0,0$ 
and $\eta,\kappa=1/2,1/2$
\begin{eqnarray*}
\eta,\kappa=0,0 &\rightarrow& \Theta(\epsilon,k)=e^{-i\frac{\pi}{D}\epsilon k}\\
\eta,\kappa=1/2,1/2 &\rightarrow& \Theta(\epsilon,k)=e^{i\frac{\pi}{2}}
e^{-i\pi \left(\epsilon-\frac{1}{2}\right) \left(\frac{k+1/2}{D}-\frac{1}{2}\right)}
\end{eqnarray*}
where $\epsilon=0,1$ and $k=0,\ldots,D-1$. The first case gives the controlled phase familiar form the QFT,
while the second is a more symmetrical interaction appropriate to the antisymmetric baker map.

\begin{figure}
\begin{center}
\includegraphics[width=0.48\textwidth]{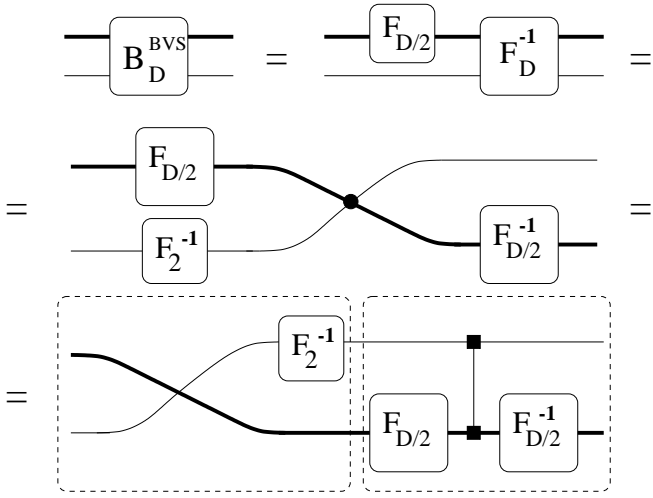}
\end{center}
\caption{\label{fig:bakersplit} Splitting of the BVS map in the product of two kernels. 
The thick line represents a space of dimension $D/2$ while the thin line is a qubit. 
The Floquet angles are implicit and the squares represent the phase interaction in 
Eq.\ref{eq:phaseint}}
\end{figure}

Using this decomposition the baker's map in the position representation can be
split into the two kernels. This is done pictorially in Fig. \ref{fig:bakersplit}. The decomposition is used
in Sec IID in the factorization of the QBM families.


\begin{thebibliography}{99}
\bibitem{Berryvoros} M. V. Berry, in {\it Chaotic Behaviour in Deterministic Systems} eds.
G. Iooss, R. Hellemann, R. Stora, North Holland, 1983, p.171.
\bibitem{schnirelman} A. J. Schnirelman, Usp. Mat. Nauk. \textbf{29}, (1974), 181; Y. Colin
de Verdiere, Comm. Math. Phys. \textbf{102} (1985)497.
\bibitem{Heller} E.J. Heller, Phys. Rev. Lett., \textbf{53}, (1984) 1515; L. Kaplan, E. Heller,
Ann. Phys.(N.Y.), \textbf{264}, 171 (1998).
\bibitem{Vergini} E. Vergini, Jour. Phys. A \textbf{33}, (2000) 4709; E. Vergini, G. Carlo,
Jour. Phys. \textbf{A33}, (2000) 4717.
\bibitem{lebouef} P. Leboeuf, A. Voros, Jour. Phys. A \textbf{23} 1765 (1990);
 S. Nonnenmacher, A. Voros, Jour. Stat. Phys. \textbf{92}, 431 (1998).
\bibitem{lakshmin3} A. Lakshminarayan and N. Meenakshisundaram, 2006 , (arXiv: nlin.CD/0603002).
\bibitem{lakshmin1} A. Lakshminarayan, J. Phys. A: Math. Gen. \textbf{38}, L597 (2005)
\bibitem{lakshmin2} N. Meenakshisundaram and A. Lakshminarayan, Phys. Rev. E \textbf{71}, 065303(R)
(2005).
\bibitem{schackcaves} 
R. Shack and M.C. Caves $2000$ {\it Applicable Algebra in Engineering,
Communication and Computing},{\bf $10$} $305$.
\bibitem{saravoros} M. Saraceno and A. Voros, 
Physica D \textbf{79} (1994) 206.
\bibitem{voros} N.L. Balazs and A. Voros, Ann. Phys, \textbf{190} (1989) 1.
\bibitem{chuang} A. Nielsen and I. Chuang, \emph{Quantum Computation and Quantum
Information}, Cambridge University Press (2000).
\bibitem{saraceno} M. Saraceno , Ann. Phys., \textbf{199} (1990) 37.
\bibitem{Coppersmith} D. Coppersmith, {\it An approximate Fourier transform useful in quantum
factoring}, IBM Research Report no. RC 19642 (1994).
\bibitem{cavesscott} A.J. Scott and C.M. Caves, 
J. Phys. A {\bf36} 9553 (2003).
\bibitem{brun} T.A. Brun, H.A. Carteret, and A. Ambainis, Phys. Rev. A \textbf{67}, 052317 (2003).
\bibitem{tanner}  G. Tanner, J. Phys. A \textbf{33}(2000) 3567.
\bibitem{nonnen1} S. Nonnenmacher and M. Zworski, 2005 (arXiv: math-ph/0505034).
\bibitem{Penrose} O. Penrose, {\it Foundations of Statistical Mechanics}, Pergamon Press, 1970.
\bibitem{scott} M.M. Tracy and A.J. Scott, J. Phys. A \textbf{35}, 8341 (2002)
\bibitem{leo} L. Ermann, J.P. Paz, M. Saraceno, Phys. Rev. A \textbf{73}, 012302 (2006).    
\bibitem{soklakov} A.N. Soklakov and R. Schack, Phys. Rev. E \textbf{61}, 5108 (2000).
\bibitem{nonnen2} N. Anantharaman and S. Nonnenmacher, 2005 (arXiv: math-ph/0512052).
\bibitem{haake} F. Haake, {\em Quantum Signatures of Chaos}, (Springer, Berlin, 1991).
\end{thebibliography}
\end{document}